\documentclass[useAMS,usenatbib]{mn2e}

\usepackage{amssymb,amsmath}
\usepackage{graphicx}
\usepackage{float}
\usepackage{multirow}
\usepackage{sidecap}
\usepackage{aas_macros}
\usepackage{fixltx2e}
\usepackage[usenames]{color}

\newcommand\msun {M$_{\odot}$}
\def\flx{erg cm$^{-2}$ s$^{-1}$}
\def\lum{erg s$^{-1}$}
\def\xmm{XMM-{\it Newton}}
\def\chan{{\it Chandra}}
\def\nftt{NGC~533}
\def\nsfo{NGC~741}
\def\am{AM~0644-741}
\def\eso{ESO~306-003}

\title[VLT/FORS2 observations of four ULX candidates]{VLT/FORS2 observations of four high-luminosity ULX candidates\thanks{Observations based on ESO programme 088.B-0076A}}
\author[M. Heida et al.]
{M. Heida$^{1,2}$, P. G. Jonker$^{1,2,3}$, M. A. P. Torres$^{1,3}$, T. P. Roberts$^4$, G. Miniutti$^5$,
\newauthor A. C. Fabian$^6$, E. M. Ratti$^1$\\
$^1$SRON Netherlands Institute for Space Research, Sorbonnelaan 2, 3584 CA Utrecht, the Netherlands\\
$^2$Department of Astrophysics/IMAPP, Radboud University Nijmegen, P.O. Box 9010, 6500 GL Nijmegen, The Netherlands\\
$^3$Harvard-Smithsonian Center for Astrophysics, 60 Garden Street, Cambridge, MA 02138, USA\\
$^4$Department of Physics, Durham University, South Road, Durham DH1 3LE, United Kingdom\\
$^5$Centro de Astrobiolog\'{i}a (CSIC--INTA), Dep. de Astrof\'{i}sica; ESAC, PO Box 78, E-28691, Villanueva de la Ca\~nada, Madrid, Spain\\
$^6$Institute of Astronomy, Madingley Road, Cambridge CB3 0HA, United Kingdom}

\begin{document}

\maketitle

\begin{abstract}
We obtained VLT/FORS2 spectra of the optical counterparts of four high-luminosity (L$_X \geq 10^{40}$ \lum) ULX candidates from the catalog of \citet{walton11b}. We first determined accurate positions for the X-ray sources from archival \chan~observations and identified counterparts in archival optical observations that are sufficiently bright for spectroscopy with an 8 meter telescope. From the spectra we determine the redshifts to the optical counterparts and emission line ratios. One of the candidate ULXs, in the spiral galaxy \eso, appears to be a bona fide ULX in an HII region. The other three sources, near the elliptical galaxies \nftt~and \nsfo~and in the ring galaxy \am, turn out to be background AGN with redshifts of 1.85, 0.88 or 1.75 and 1.40 respectively. Our findings confirm the trend of a high probability of finding background AGN for systems with a ratio of log(F$_X$/F$_{\textrm{opt}}$) in the range of -1 -- 1.
\end{abstract}


\section{Introduction}
Ultraluminous X-ray sources (ULXs) are off-nuclear X-ray sources in galaxies with an X-ray luminosity above the Eddington luminosity of a 10 \msun~black hole, or $\sim 10^{39}$ \lum. Several scenarios have been proposed to explain their high luminosities. Geometrical \citep{king01} or relativistic \citep{kording02} beaming may allow for the observation of super-Eddington luminosities. In some sources there is evidence for a new state with truly super-Eddington accretion rates \citep{gladstone09}. Recent investigations into the X-ray luminosity function (XLF) of ULXs suggest that the majority of ULXs are formed by the high-luminosity tail of X-ray binaries (XRBs) and contain stellar mass black holes (BHs) \citep{swartz11, mineo12}. The best-fitting XLF exhibits a cut-off around $10^{40}$ \lum, suggesting that this may be the effective upper limit for the luminosity of the most massive objects in the sample.

However, \citet{swartz11} argue that if they extrapolate their best-fitting XLF, based on a complete sample of ULXs within 14.5 Mpc, to larger distances, they can not explain the relatively large number of ULXs with luminosities above $10^{41}$ \lum~that have been observed. Hence these ULXs may belong to a different class of objects. These sources would have to exceed the Eddington limit by more than a factor 100 if they contained stellar-mass black holes. They may be good candidates to host the predicted but thus far elusive intermediate-mass black holes (IMBHs). These IMBHs may form in the collapse of a dense stellar cluster \citep{portegieszwart02}, the collapse of population III stars in the early universe \citep{madau01} or the direct collapse of massive gas clouds \citep{begelman06}. They may reside in globular clusters \citep{maccarone07}, but conclusive evidence for their existence there has not yet been found. For a review on IMBHs and their formation mechanisms see \citet{vandermarel04}. The best candidate for an IMBH to date is the extremely bright source HLX-1 in ESO 243-49, which reaches maximum X-ray luminosities of $\sim 10^{42}$ \lum \citep{farrell09}. The recent work by \citet{sutton12} provides more evidence for extreme ULXs as IMBHs.

Most ULX candidates are discovered by searching for off-nuclear X-ray point sources in galaxies (e.g. from the \chan~or \xmm~serendipitous source catalogs; see for example \citealt{walton11b, liu11}). The ULX catalogs compiled in this way are contaminated with objects that also show up as off-nuclear, bright X-ray sources but are not accreting IMBHs or stellar-mass BHs. Background active galactic nuclei (AGN) and quasars are obvious examples, but some X-ray bright supernovae (most likely type IIn, \citealt{immler03}) and active foreground stars may also contaminate the catalogs. One way to identify these contaminants if the ULX candidate has a bright optical counterpart is to take an optical spectrum. If emission or absorption lines are present the redshift to the source can be measured. In this way we can determine whether the source is associated with the galaxy or is a background or foreground object (compare e.g. \citealt{gutierrez13}).

If the X-ray source is associated with the galaxy, optical spectra can give us additional information to classify the object. Some ULXs \citep{pakull02, kaaret09} are surrounded by bubbles of ionized gas, which can act as calorimeters and as such tell us if the emission is strongly beamed or not. The intensity ratios of the emission lines from these regions provide information on the source of the ionizing radiation, e.g. whether they are shock ionized or X-ray photo-ionized (e.g. \citealt{abolmasov07}).

We selected four high-luminosity (L$_X \geq 10^{40}$ \lum) ULX candidates from the catalog of \citet{walton11b} with accurate positions that we measured using archival \chan~observations and optical counterparts that are sufficiently bright for optical spectroscopy. Two of the ULX candidates are situated in elliptical galaxies (\nftt~and \nsfo). ULX candidates in elliptical galaxies have a higher chance to be background AGN (39\%, compared to 24\% for all sources in the catalog of \citet{walton11b}). On the other hand, IMBHs may form in dense (globular) clusters \citep{portegieszwart02, miller02} and the optical counterparts to these ULX candidates could well be just that, making them interesting targets for further investigation. \am~is a ring galaxy with a ULX candidate situated in between the nucleus and the ring. \eso~is a spiral galaxy with a ULX in the outer edge of the disk, apparently associated with an extended optical source. We obtained optical spectra of these four sources with the FOcal Reducer and low dispersion Spectrograph (FORS2) mounted on the Very Large Telescope (VLT) \citep{appenzeller98}. The observations and data reduction steps are described in Section 2; Section 3 contains the results. In Section 4 we discuss our findings.

\section{Observations and data reduction}
\subsection{X-ray observations}
We use archival \chan~observations to get exact source positions for the ULX candidates in \nftt, \nsfo, \am~and \eso. Table \ref{chantab} lists the details of all observations. 

\begin{table*}
\begin{minipage}{135mm}
\caption{The \textit{Chandra} observations of the four ULX candidates.}\label{chantab}
 \begin{tabular}{lccccc}
 \hline
 Galaxy & Observation ID & Exposure time & Source on CCD & Off-axis angle & Observation date \\
 & & (kiloseconds)& & (arcmin) & (UT)\\
 \hline
\nftt~& 2880 & 38.1 & ACIS S3 & 0.85 & 2002-07-28\\
 \nsfo~& 2223 & 30.74 & ACIS S3 & 2.74 & 2001-01-28\\
 \am~& 3969 & 39.97 & ACIS S3 & 0.57 & 2003-11-17\\
 \eso~& 4994 & 22.75 & ACIS I3 & 6.60 & 2004-03-10\\
 \hline
 \end{tabular}
 
\medskip
Notes: \chan~observation ID number, exposure time in kiloseconds, CCD on which the source was detected, the off-axis angle of the source in arcminutes and the observation date.
\end{minipage}
\end{table*}

We use \textsc{Ciao} version 4.4 to process the \chan~observations, with the calibration files from \textsc{caldb} version 4.5.0. We treat the \chan~observations as follows: first we update the event files with ACIS\_{}PROCESS\_{}EVENTS, then we use WAVDETECT to find the position of the ULX candidate. Sources within 3 arcmins of one of the ACIS aimpoints have a 90\% confidence error circle around the absolute position with a radius of 0.6''; this is valid for the ULX candidates in \nftt, \nsfo~and \am. The candidate in \eso~has 25 counts and was observed at 6.6' off-axis, which means it has a 95\% confidence error circle with a radius of $\sim$1.5'' \citep{hong05}. For the sources in \nftt, \nsfo~and \am~we extract the source counts in a circle with 6 pixel radius (90\% encircled energy fraction) around the source positions using SPECEXTRACT. For the ULX candidate in \eso~we use a circle with a radius of 10 pixels to get the same encircled energy fraction, since it was observed at 6.6' off-axis. As background regions we use circles with 80 pixel radius on the same CCD but not containing any sources. We use \textsc{XSpec} version 12.6.0 to fit an absorbed powerlaw (pegpwrlw) to the data in the 0.3-8 keV range. We then extrapolate to get the 0.2-12 keV flux to compare this with the values reported by \citet{walton11b}. For consistency we adopt the same model parameters: a photon index of 1.7 and $N_H = 3 \times 10^{20}$ cm$^{-2}$, and allow only the flux to vary. We find that all \textit{Chandra} fluxes are consistent with those from \xmm~as reported by \citet{walton11b}. The positions of the X-ray sources and their fluxes are summarized in Table \ref{chansrc}.

\begin{table*}
\begin{minipage}{90mm}
\caption{The positions and unabsorbed 0.2-12 keV X-ray fluxes of the ULX candidates.}\label{chansrc}
 \begin{tabular}{ccccc}
 \hline
 Host galaxy & Right Ascension & Declination & Source flux \\
 &&& (\flx)\\
 \hline
\nftt~& 01:25:33.63 & +01:46:42.6& $2.9 \pm 0.2 \times 10^{-14}$ \\
 \nsfo~& 01:56:16.14 & +05:38:13.2 & $2.5 \pm 0.3 \times 10^{-14}$ \\
 \am~& 06:43:02.24 & -74:14:11.1 & $3.5 \pm 0.2 \times 10^{-14}$ \\
 \eso~& 05:29:07.21 & -39:24:58.4 & $2.4 \pm 0.4 \times 10^{-14}$ \\
 \hline
 \end{tabular}
 
\medskip
Notes: the positions of the ULX candidates in \nftt, \nsfo~and \am~are accurate to within 0.6''(90\% confidence level), for the source in \eso~this value is 1.3''. We fit the fluxes assuming an absorbed powerlaw with photon index 1.7 and $N_H = 3 \times 10^{20}$ cm$^{-2}$ for consistency with the method used by \citet{walton11b}. 
\end{minipage}
\end{table*}

\subsection{Optical images and photometry}
To find the optical counterparts of the ULX candidates we use archival optical observations of their host galaxies. \nftt~and \nsfo~were observed as part of the Sloan Digital Sky Survey (SDSS), and we use the SDSS $r'$-band images to identify the optical counterparts to the ULX candidates in these galaxies (Figure \ref{fig:agn}). There is no photometric information for the source in \nftt, so we use the aperture photometry tool in \textsc{GAIA} to estimate the $r'$-band magnitude. SDSS does provide \textit{u'}, \textit{g'}, \textit{r'}, \textit{i'} and \textit{z'} magnitudes for the object in \nsfo, but these are incorrect because the source is too close to the edge of the frame. Therefore we also use \textsc{GAIA} to estimate the $r'$-band magnitude for this source. For both optical counterparts we find that $r' = 21 \pm 1$.

The Hubble Space Telescope (HST) archive contains several observations of \am~made with the Advanced Camera for Surveys (ACS). We use the V-band (F555W) image with exposure identifier j8my05o2q, observed on 2004-01-16 with an exposure time of 2200 seconds (see Figure \ref{fig:agn}). We visually compare the position of point sources from the USNO CCD Astrograph Catalog (UCAC) 3 \citep{zacharias09} with their counterparts in the HST image and find that the astrometric calibration of the image does not need to be improved. The ULX candidate has a counterpart that is in the DAOPHOT source list of this HST image in the Hubble Legacy Archive (HLA\footnote{http://hla.stsci.edu}). It has a V-band magnitude of $21.79 \pm 0.05$.

We identify the optical counterpart to the ULX candidate in \eso~in a 480 seconds R-band observation made on 2004-01-25 UT with VLT/VIMOS that we retrieved from the ESO archive. Its R-band magnitude is approximately 21, with the caveat that this is an extended source in a region with a very high background level due to the galaxy, which means that this measurement is not very accurate. We also obtained a \textit{g'}-band, 120 seconds exposure of this galaxy in our VLT/FORS2 run (see Figure \ref{fig:eso306}), of which we visually inspected the astrometric solution by comparing the positions of bright stars with those in the UCAC 3.

\subsection{Optical spectroscopy}
We obtained VLT/FORS2 observations of \nsfo~($3 \times 1800$ s), \am~($3 \times 1800$ s) and \eso~($2 \times 2700$ s) on 2011-12-03 UT under programme 088.B-0076A using the GRIS\_600V grism and a 1\arcsec~slit width. This configuration covers the wavelength range 4430-7370 \AA{}~with a dispersion of 0.74 \AA{}/pixel,  yielding a resolution of 4.25 \AA{}~for the 1'' slit (measured at 6300 \AA). This allows us to observe the H$\alpha$, [N{\sc II}] complex and the H$\beta$ and [O{\sc III}] lines if the sources are located at the same distance as their apparent host galaxies, with high enough resolution to separate them. The night was photometric so we also observed several spectrophotometric standard stars to perform a flux calibration. The seeing varied between 0.7 and 1.1\arcsec. %peaked around 05:00 UT
The spectra of \nftt~($3 \times 1500$ s) were made in service mode on 2012-01-16 UT with the GRIS\_300V+10 grism and a 0.5\arcsec~slit width, giving a wavelength coverage from 4450-8700 \AA{}~with a dispersion of 1.68 \AA{}/pixel and a spectral resolution of 6.4 \AA{}~for the 0.5'' slit (measured at 6300 \AA{}). The seeing varied during the night and we have no observations of spectrophotometric standards.

To reduce the spectra we use the \textsc{starlink} software package \textsc{Figaro} and the \textsc{Pamela} package developed by Tom Marsh\footnote{http://deneb.astro.warwick.ac.uk/phsaap/software}. We follow the steps outlined in the \textsc{Pamela} manual to extract the spectra, using Keith Horne's optimal extraction algorithm \citep{horne86}. We then use the software package \textsc{Molly}, also by Tom Marsh\footnotemark[\value{footnote}], to perform the wavelength calibration and, for the data taken on 2011-12-03, the flux calibration. We do not correct for telluric absorption. Because we have multiple spectra of each source we average them to get a better signal-to-noise ratio. The two observations of \eso~were taken under varying seeing conditions. Because of this the continuum level is different in the two spectra, so we normalize these spectra before averaging them.
We use \textsc{Molly}'s MGFIT task to fit Gaussian profiles to the emission lines in the spectra to determine the full width at half maximum (FWHM) of the lines and the redshift to the sources. 

%The flux calibration is not absolute because we cannot account for slit losses, but it does correct for the wavelength-dependent instrumental response.

\section{Results}

\begin{table*}
\begin{minipage}{150mm}
\caption{Source properties of the background AGN}\label{optinfo}
 \begin{tabular}{lccccc}
 \hline
 Source name & In galaxy & z & Line & FWHM & Log(F$_X$/F$_{\textrm{opt}}$) \\
 &&&& km/s &\\
 \hline
CXOU J012533.3+014642 & \nftt~& $1.8549 \pm 0.0003$ & C\textsc{IV} & $2300 \pm 70 $ & $0.0 \pm 0.5$\\
& & & C\textsc{III}] & $7800 \pm 200$ & \\
& & & Mg{\sc II} & $5700 \pm 200$ & \\
CXOU J015616.1+053813 & \nsfo~& $0.8786 \pm 0.0006$ or & Mg{\sc II} or & $8400 \pm 200$ & $0.0 \pm 0.5$\\
& & $1.7535 \pm 0.0009$ & C{\sc III}] & & \\
CXO J064302.2-741411 & \am~& $1.3993 \pm 0.0001$ & C{\sc III}] & $5100 \pm 70$ &  $0.7 \pm 0.1$\\
 & & & Mg{\sc II} & $4220 \pm 40$ &\\
 \hline
 \end{tabular}
 
\medskip
Notes: Lines used for the redshift determination to the quasars, their FWHM in km/s and the X-ray to optical flux ratio of these sources. The X-ray to optical flux ratios are calculated using the \xmm~0.2-12 keV fluxes from \citet{walton11b} and the \textit{r'}-band (for \nftt~and \nsfo) or V-band (for \am) optical fluxes.
\end{minipage}
\end{table*}

\begin{figure*}
\hbox{
\includegraphics[width=0.33\textwidth]{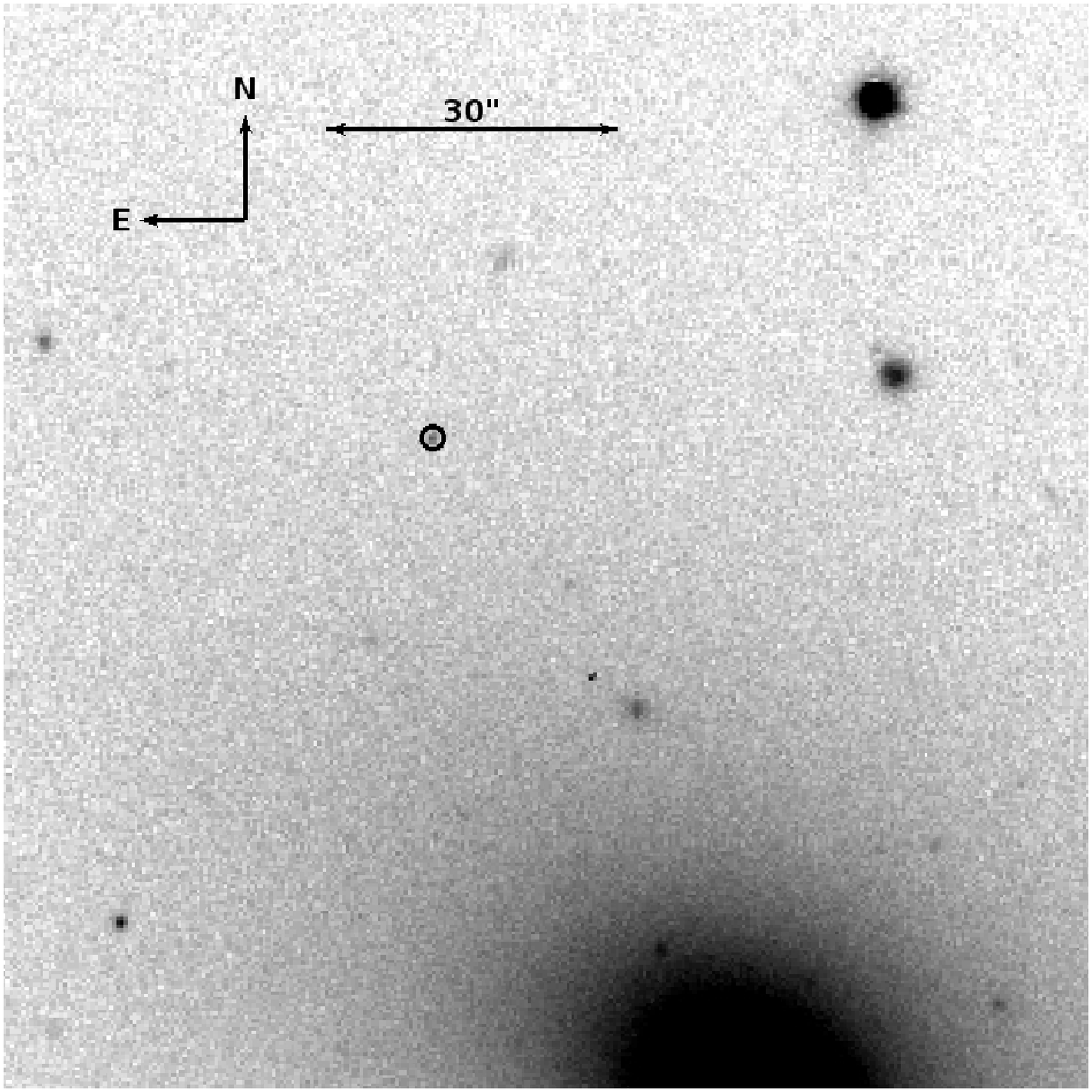}
\includegraphics[width=0.33\textwidth]{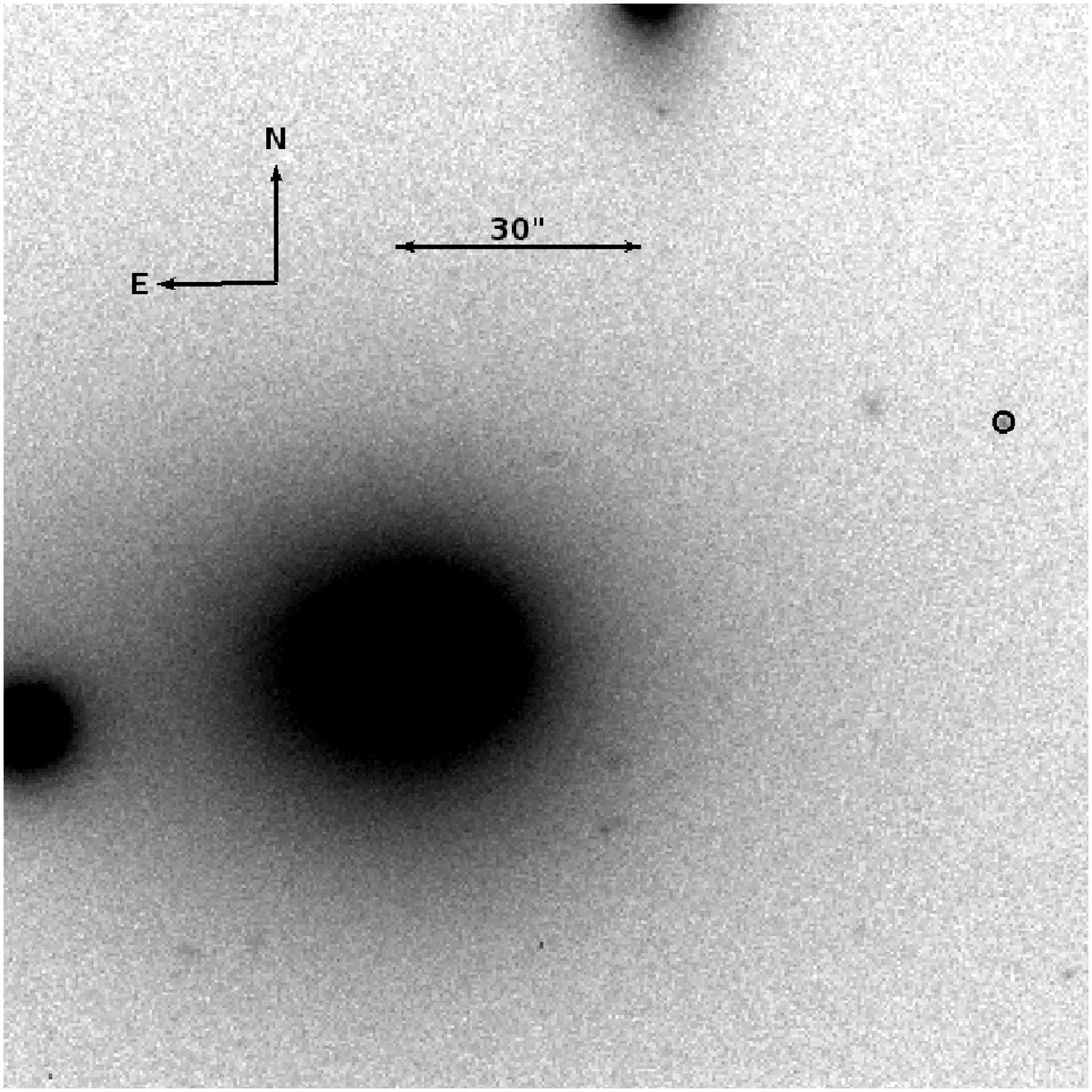}
\includegraphics[width=0.33\textwidth]{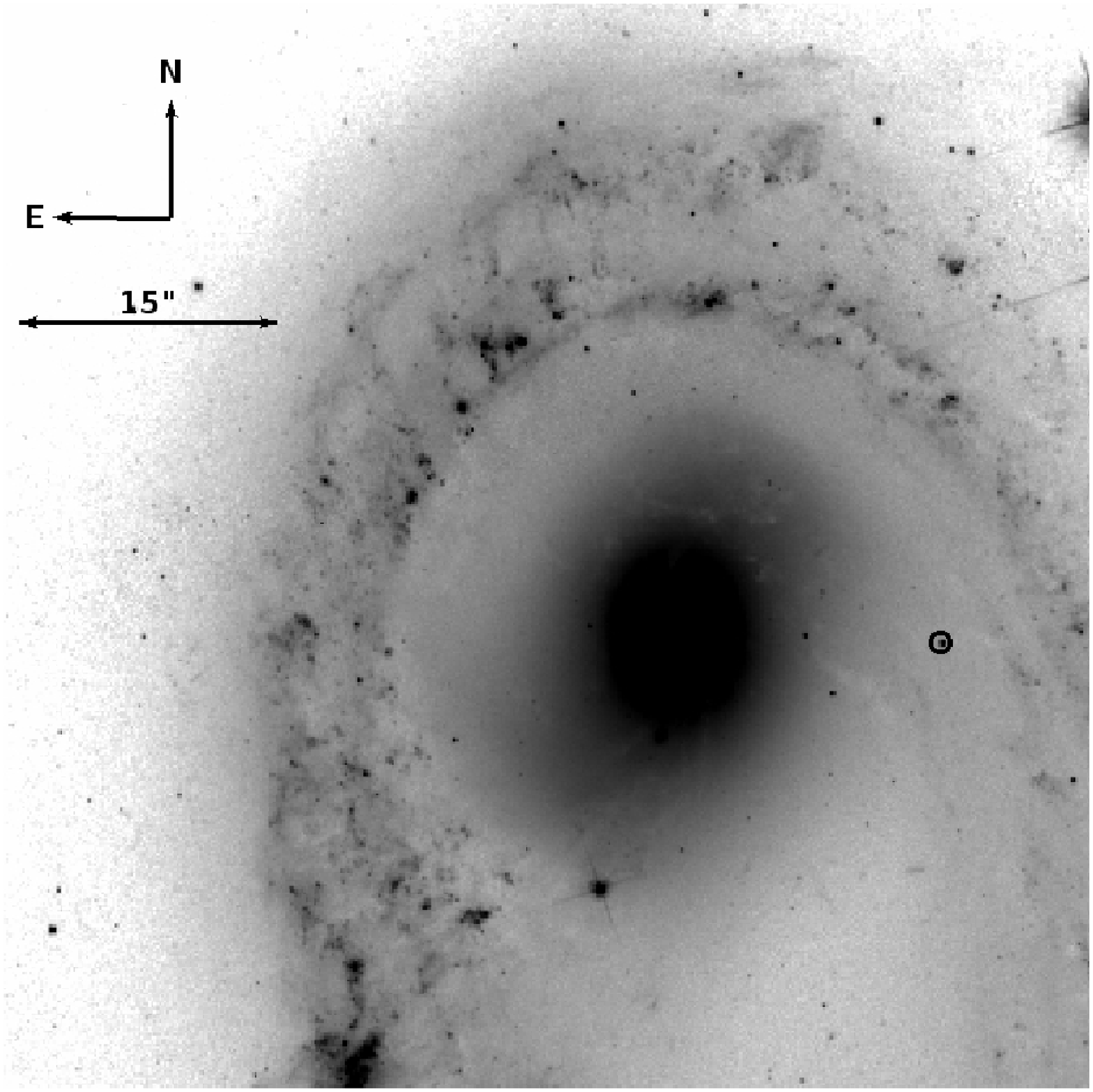}
}
\hbox{
\includegraphics[width=0.33\textwidth]{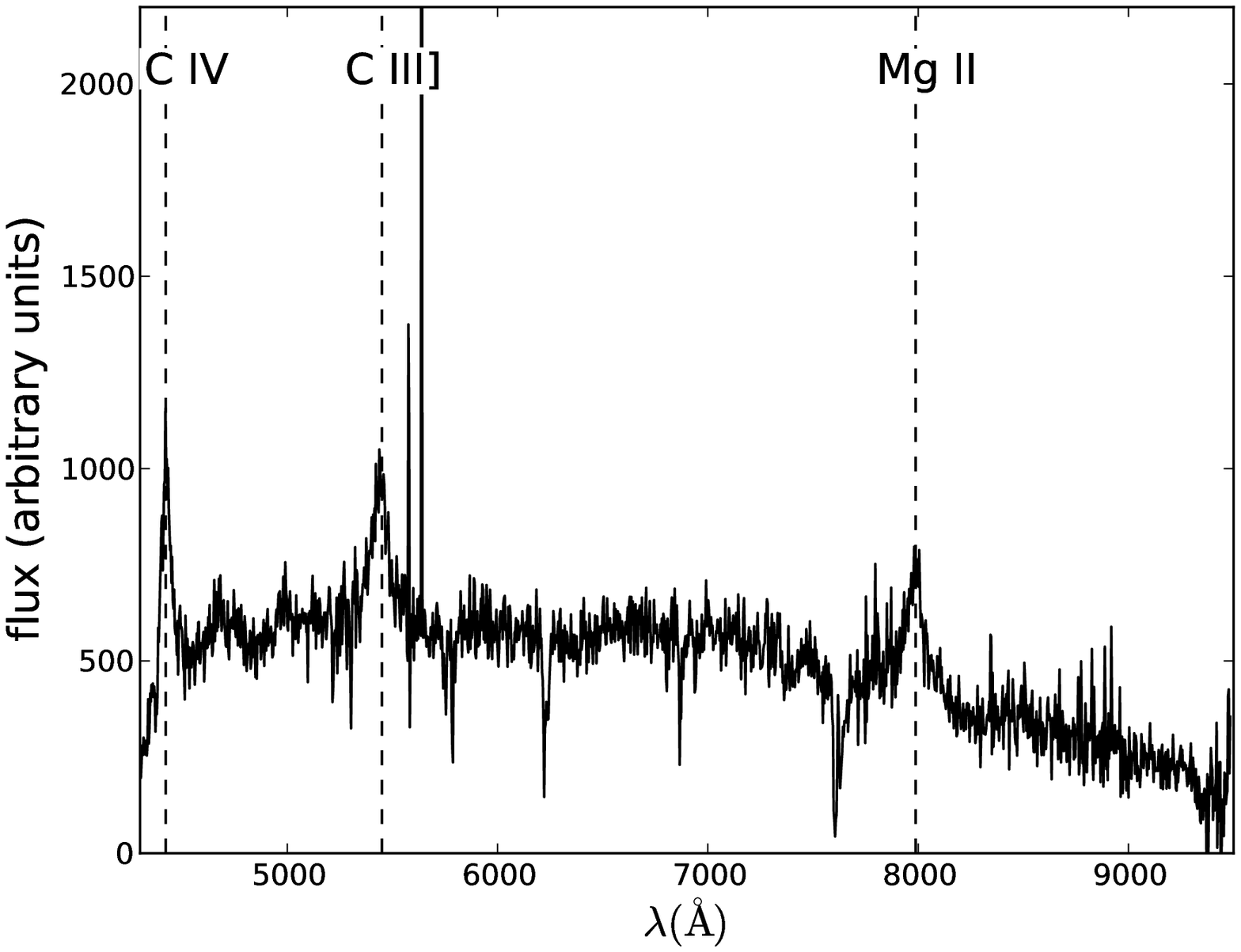}
\includegraphics[width=0.33\textwidth]{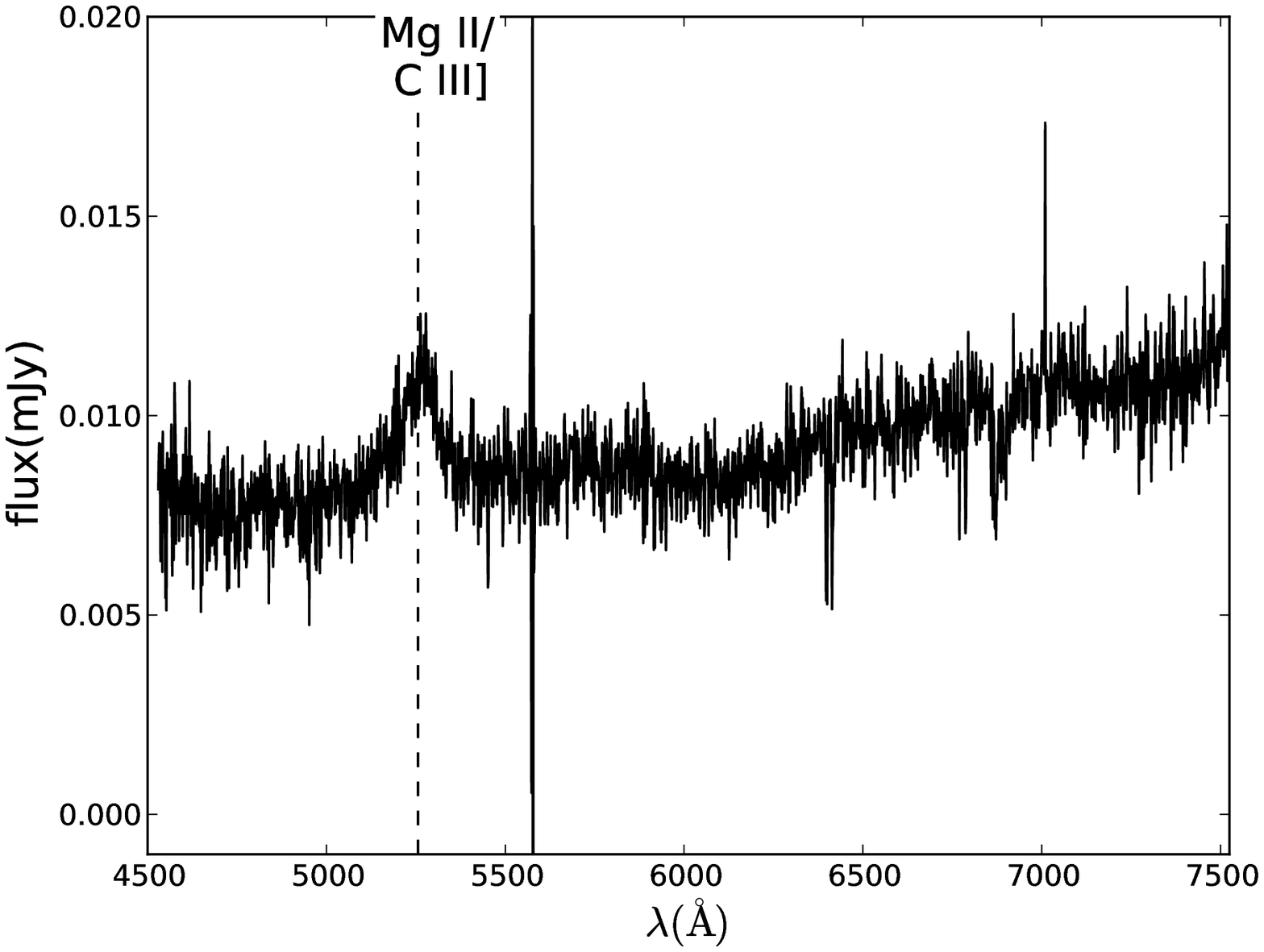}
\includegraphics[width=0.33\textwidth]{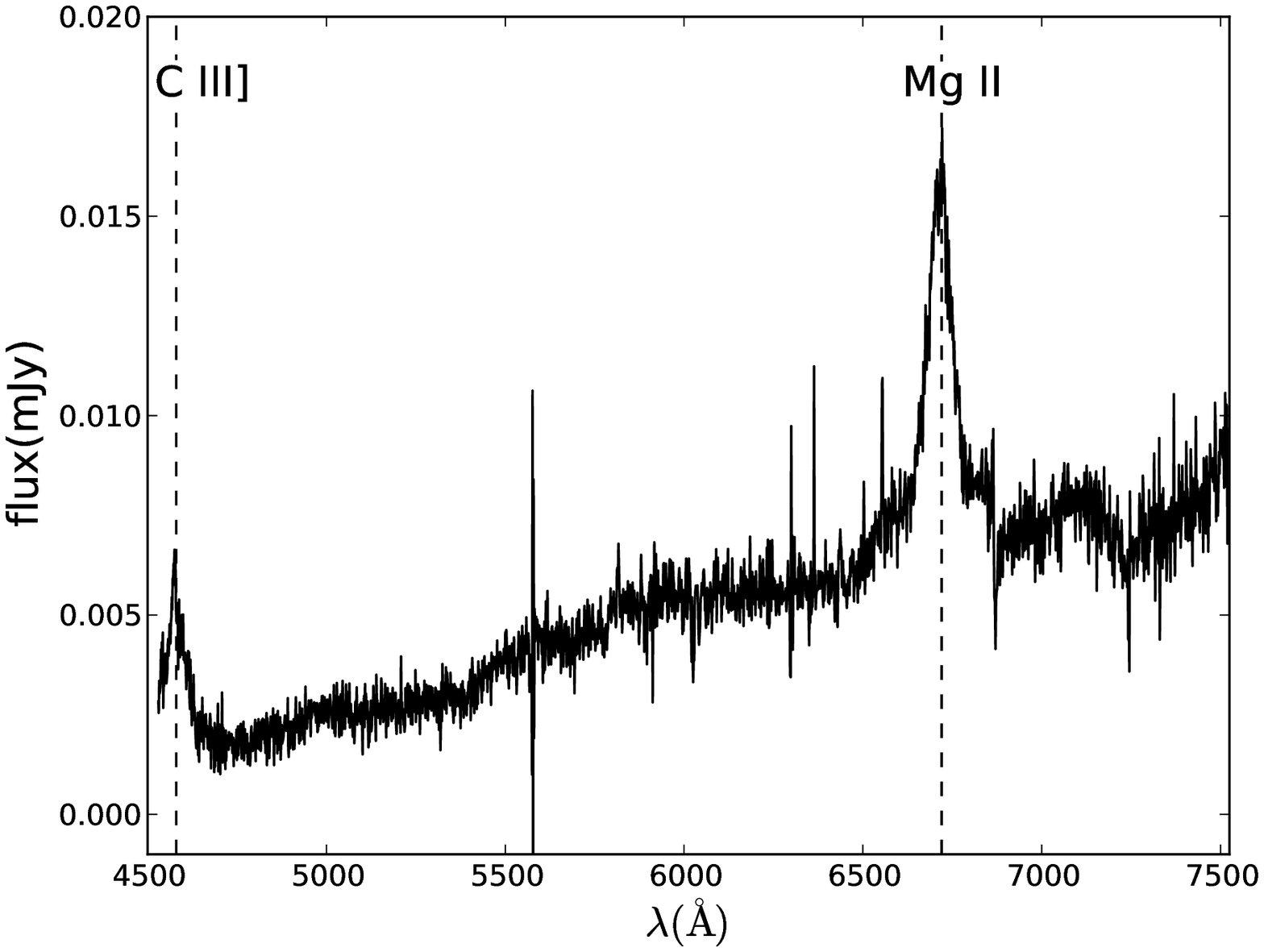}
}
\caption{The finders and FORS2 spectra of the three ULX candidates that are background AGN. The 90\% confidence error circles around the X-ray positions have a radius of 0.6'', for \nftt~and \nsfo~we plot a larger circle for visual clarity. \emph{Left:} The SDSS \textit{r'}-band image of \nftt~with a 1.2'' radius circle around the \chan~position of the ULX candidate and the spectrum in which the C\textsc{IV}, C{\sc III}] and Mg{\sc II} emission lines, redshifted by $z=1.85$, are marked. The absorption features at 6200 \AA{}~are caused by interstellar absorption, and those at 6900 \AA{}~and 7600 \AA{}~are telluric in origin. \emph{Middle:} The SDSS \textit{r'}-band image of \nsfo~with a 1.2'' radius circle around the \chan~position of the ULX candidate and the spectrum of the optical counterpart. The marked emission line can be either Mg{\sc II} $\lambda2798$ line redshifted by $z=0.88$ or C{\sc III}] at $z = 1.75$. \emph{Right:} An HST ACS V-band image of \am~with the 0.6'' radius error circle around the \chan~position of the ULX candidate, and the spectrum with the Mg{\sc II} and C{\sc III}] lines, redshifted by $z=1.40$, marked.}\label{fig:agn}
\end{figure*}

\subsection{\nftt}
\nftt~is the dominant elliptical galaxy in a group with the same name at $z = 0.0185$ \citep{smith00}. The ULX candidate is located at 78'' from the center of the galaxy that has a semi-minor axis of 90'' (based on the D25 isophote, \citealt{nilson73}). The X-ray source has an unresolved optical counterpart that is visible in the image of the SDSS, with \textit{r'}-band magnitude $\approx 21$. Figure \ref{fig:agn} shows the galaxy with the position of the ULX candidate and the FORS2 spectrum of the source. 

Three broad emission lines are visible. We identify these as C\textsc{IV}, C{\sc III}] and Mg{\sc II} at $z = 1.8549 \pm 0.0003$. This proves the ULX candidate to be a background AGN, not associated with \nftt. The 0.2-12 keV X-ray luminosity calculated for this source by \citet{walton11b} was $(2 \pm 1) \times 10^{40}$ \lum, assuming a distance to the ULX of 73.8 Mpc. The true distance to this source is 4730 Mpc (using $H_0 = 75$ km/s/Mpc for consistency with \citealt{walton11b}), which gives this AGN an X-ray luminosity of $(7 \pm 4) \times 10^{43}$ \lum~using the flux as measured with \xmm.

%\begin{figure*}
%\hbox{
%\includegraphics[width=0.4\textwidth]{ngc533_sdss}
%\includegraphics[width=0.6\textwidth]{ngc533_avspec}
%}
%\caption{\emph{Left:} The SDSS \textit{r'}-band image of \nftt~with a 1.2'' radius circle around the \chan~position of the ULX candidate. The 90\% confidence error circle around the position has a radius of 0.6'', we plot a larger circle for visual clarity. \emph{Right:} The FORS2 spectrum of the optical counterpart. The C\textsc{IV}, C{\sc III}] and Mg{\sc II} emission lines, redshifted by $z=1.85$, are marked. The absorption features at 6200 \AA{}~are caused by interstellar absorption, and those at 6900 \AA{}~and 7600 \AA{}~are telluric in origin.}\label{fig:ngc533}
%\end{figure*}

\subsection{\nsfo}
\nsfo~is an elliptical galaxy located at $z = 0.0185$ with a (D25) semi-major axis of 92.7'' \citep{devaucouleurs91}. The ULX candidate is located 78'' West of the center of \nsfo~and has a counterpart that is unresolved in the SDSS image. Its \textit{r'}-band magnitude is $\sim 21$. Figure \ref{fig:agn} shows the SDSS \textit{r'}-band image of \nsfo~with the position of the ULX candidate and the FORS2 spectrum of the counterpart.

The spectrum shows one broad emission line, with a FWHM of 147 \AA{}. We cannot say with certainty which line this is. The most likely options are that it is either the Mg{\sc II} $\lambda2798$ line or the C{\sc III}] $\lambda1909$ line. In the first case this ULX candidate would be a background AGN at a redshift of $z = 0.8786 \pm 0.0006$ with an X-ray luminosity of $(1.6 \pm 0.6) \times 10^{43}$ \lum. In the second case it would be at $z = 1.7535 \pm 0.0009$, with $L_X = (4.2 \pm 1.5) \times 10^{43}$ \lum. In both cases the source is not a ULX but a background AGN, unconnected to \nsfo.
%\begin{figure*}
%\hbox{
%\includegraphics[width=0.4\textwidth]{ngc741_sdss}
%\includegraphics[width=0.6\textwidth]{ngc741_avspec}
%}
%\caption{\emph{Left:} The SDSS \textit{r'}-band image of \nsfo~with a 1.2'' radius circle around the \chan~position of the ULX candidate. The 90\% confidence error circle around the position has a radius of 0.6'', we plot a larger circle for visual clarity. \emph{Right:} The FORS2 spectrum of the optical counterpart. The marked emission line can be either Mg{\sc II} $\lambda2798$ line redshifted by $z=0.88$ or C{\sc III}] at $z = 1.75$.}\label{fig:ngc741}
%\end{figure*}

\subsection{\am}
\am~is a ring galaxy at $z = 0.022$ that shows signs of recent interaction with a smaller galaxy \citep{few82, lauberts89}. The ULX candidate in this galaxy is located in between the core of the galaxy and the ring. A point-like object with a V-band magnitude of 21.8 is visible at the position of the X-ray source in archival HST images (see Figure \ref{fig:agn}).

The FORS2 spectrum of the counterpart shows two emission lines that we identify as Mg{\sc II} and C{\sc III}] at $z=1.3993 \pm 0.0001$. This ULX candidate is another background AGN with an 0.2-12 keV luminosity of $(8.1 \pm 0.8) \times 10^{43}$ \lum.

%\begin{figure*}
%\hbox{
%\includegraphics[width=0.4\textwidth]{a0644_hst}
%\includegraphics[width=0.6\textwidth]{a0644_avspec}
%}
%\caption{\emph{Left:} An HST ACS V-band image of \am~with the 0.6'' radius (90\% confidence) error circle around the \chan~position of the ULX candidate. \emph{Right:} The FORS2 spectrum of the optical counterpart. The Mg{\sc II} and C{\sc III}] lines, redshifted by $z=1.40$, are marked.}\label{fig:a0644}
%\end{figure*}

\subsection{\eso}
\begin{figure*}
\hbox{
\includegraphics[width=0.4\textwidth]{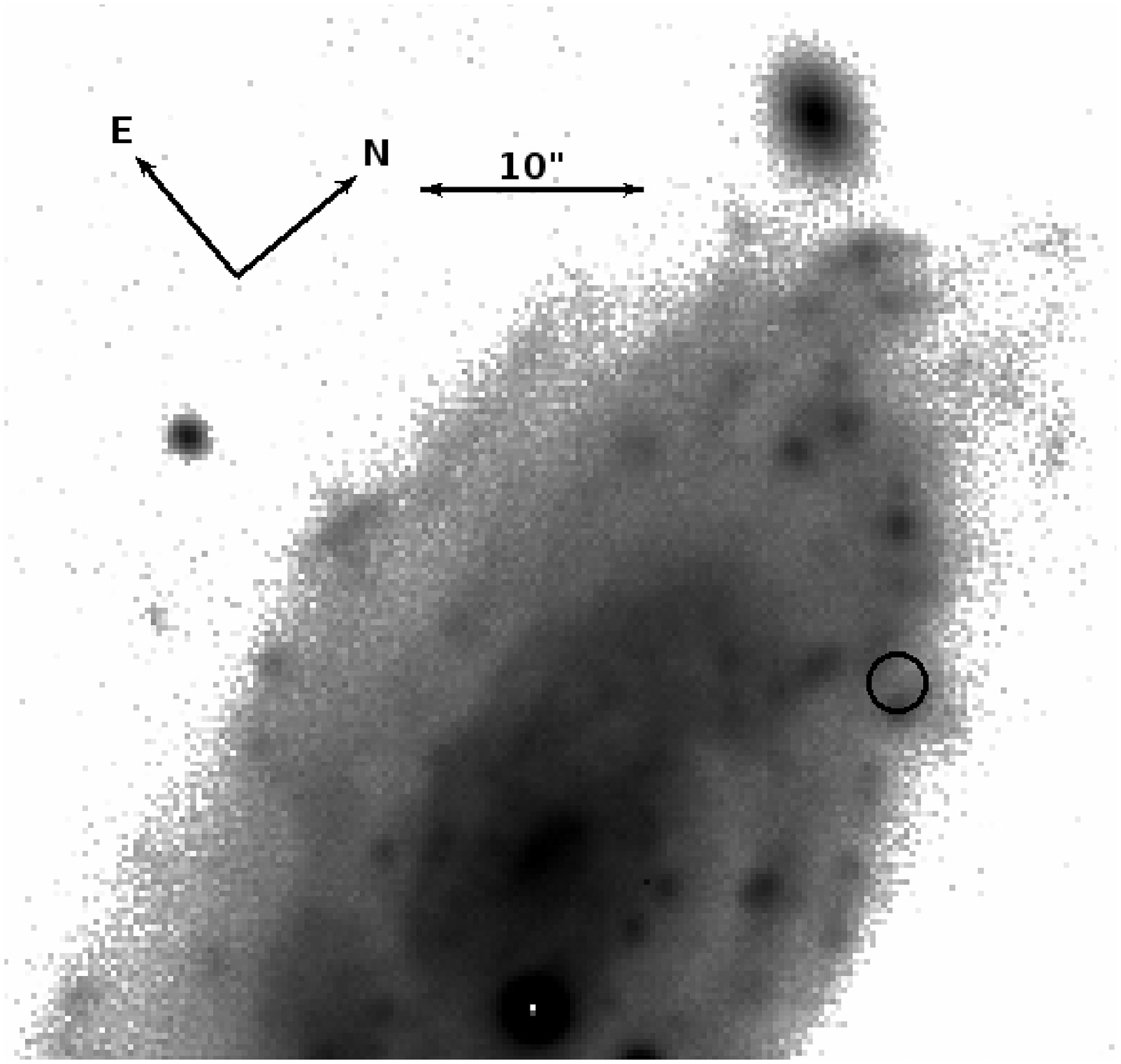}
\includegraphics[width=0.6\textwidth]{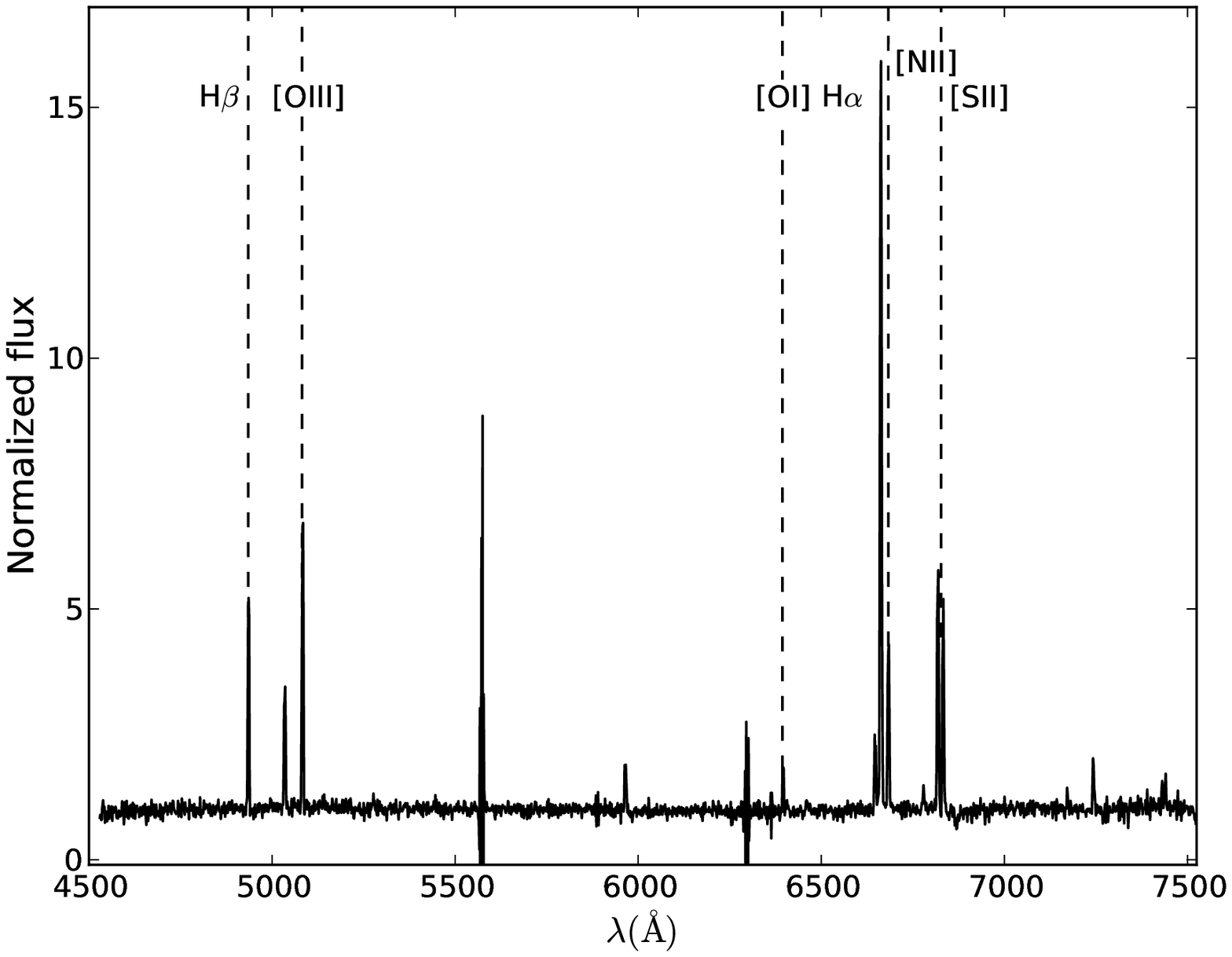}
}
\caption{\emph{Left:} The FORS2 \textit{g'}-band acquisition image of \eso~with the 1.3'' radius (90\% confidence) error circle around the position of the ULX candidate. \emph{Right:} The FORS2 spectrum of the candidate optical counterpart to the X-ray source. Several emission lines, redshifted by $z=0.016$, are marked.}\label{fig:eso306}
\end{figure*}

The spiral galaxy \eso, at $z \approx 0.016$ \citep{dasilva06}, contains a ULX candidate that is located on the edge of the spiral structure (see Figure \ref{fig:eso306}). An optical source is visible on the edge of the error circle. Visual inspection shows the profile of the counterpart to be  more extended than that of point sources in the same image, but because of the high background level and steep gradient it is not possible to perform an acceptable fit to the profile. The full width at half maximum (FWHM) of point sources in this image (provided by the seeing) is 0.8''. At the distance of \eso~this yields a lower limit to the size of the source of 240 pc. The two spectra that we obtained of this source show slightly different line ratios and continuum levels (for example, the H$\beta$/H$\alpha$ ratio changes by 10\%). This can be explained by seeing variations if the optical counterpart to this source is extended: then slit losses can cause the small changes in the line ratios if there are intrinsic spatial variations in the line ratios in the extended source.

\begin{figure}
\includegraphics[width=0.5\textwidth]{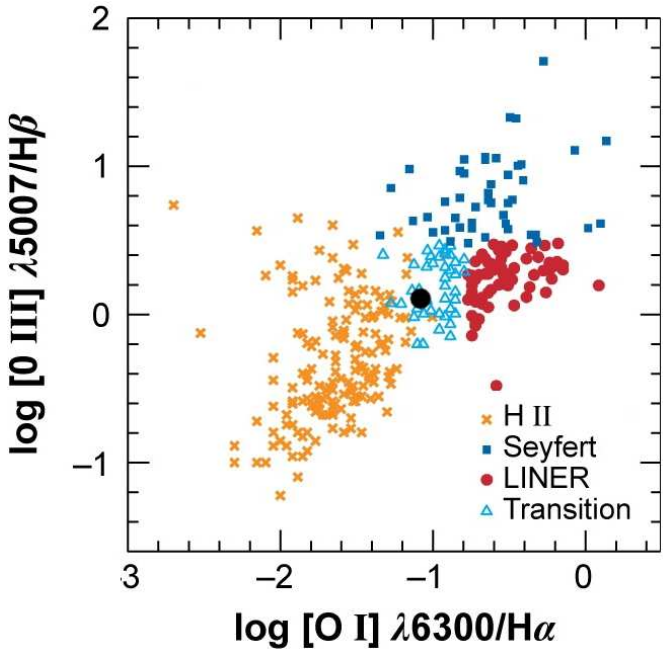}
\caption{[O{\sc I}] $\lambda$6300/H$\alpha$ versus [O{\sc III}] $\lambda$5007/H$\beta$ line ratios for H{\sc II} regions, AGN (LINERs %(low-ionization nuclear emission-line regions)
and Seyferts) and transition objects (figure adapted from \citealt{ho08}). The black dot represents the line ratios for the ULX in \eso.}\label{fig:dd}
\end{figure}

The spectrum is similar to that of an H{\sc II} region, with narrow Hydrogen emission lines and strong forbidden lines. The redshift of the lines equals that of the center of the galaxy, indicating that if the X-rays are associated with this optical source, this is a bona fide ULX with a luminosity of $1.4 \pm 0.3 \times 10^{40}$ \lum~based on the \xmm~flux measured by \citet{walton11b}. The X-ray flux is constant between the \xmm~and \chan~observations. The X-ray to optical flux ratio of the source is log(F$_X$/F$_{\textrm{opt}}$)$ = 0.3 \pm 0.5$, based on the \xmm~0.2-12 keV flux from \citet{walton11b} and the \textit{r'}-band flux. The line ratios, especially the [O{\sc I}] $\lambda$6300/H$\alpha$ ratio, place the source among the transition objects in the diagnostic diagrams of \citet{ho08} (see Figure \ref{fig:dd}). The He{\sc II} $\lambda4686$ emission line has been detected in several ULX nebulae \citep{pakull02, kaaret09}, but we do not detect it here, possibly because the sensitivity of the detector drops off steeply towards the blue end. The 2-$\sigma$ upper limit for the equivalent width of this line is 1.0 \AA{}. This corresponds to a flux of $\sim 10^{-17}$ \flx or a luminosity of $\sim 5 \times 10^{36}$ \lum.

\section{Discussion}
We obtained VLT/FORS2 spectra of the optical counterparts of four bright ULX candidates with accurate positions obtained by us from archival \chan~observations. Two of these are located in elliptical galaxies \nftt~and \nsfo. Another candidate is situated in \am, a ring galaxy that recently interacted with a small elliptical galaxy, and in the spiral galaxy \eso. Three of our four targets turn out to be background AGN with X-ray luminosities ranging from 1 to 8 $\times 10^{43}$ \lum; one (in \eso) seems to be a bona fide ULX. 

The fraction of background AGN in our sample is higher than the fraction estimated by \citet{walton11b} for their catalog. Although this can be due to small number statistics since we only investigate four sources, it is in line with results from other spectroscopic studies of ULX candidates. Optical spectroscopy of a sample of 23 ULX candidates in total yielded 20 background AGN and three foreground stars (\citealt{gutierrez05, gutierrez06, gutierrez07, gutierrez13}). Another study that targeted 17 ULX candidates from the catalog of \citet{colbert02} found that 15 were background AGN and the other two objects were foreground stars \citet{wong08}.

All these studies mainly target ULX candidates that are relatively isolated and have a bright optical counterpart, a selection effect induced by the relative ease with which spectroscopic observations can be carried out for these sources. Sources located in crowded areas, like the spiral arms of late type galaxies, are more difficult targets for ground based optical spectroscopic observations. This means that spectroscopic studies are aimed at ULX candidates that have a low X-ray to optical flux ratio and that are situated relatively far away from their suspected host galaxies. As the authors of these previous papers also note, these selection criteria introduce a bias towards background AGN.

A possible method to select ULX candidates that are most likely to be real ULXs is to calculate the expected contribution of background sources based on the known density of AGN in X-ray and optical observations (\citealt{lopez06, sutton12}). Alternatively it may be possible to use the X-ray to optical flux ratios of ULX candidates to select targets for future spectroscopic studies. All our sources have X-ray to optical flux ratios log(F$_X$/F$_{\textrm{opt}}$) in the range between -1 and 1, typical for AGN (e.g. \citealt{barger03}). Most ULXs show values for log(F$_X$/F$_{\textrm{opt}}$) ranging from 2 - 3 (\citealt{tao11, tao12, sutton12}). The low value that we find for the ULX in \eso~can be explained if we assume that we do not resolve the ULX counterpart but instead observe the optical flux of the entire HII region that it resides in, thus lowering log(F$_X$/F$_{\textrm{opt}}$).

However, if we were to select candidates for spectroscopy on the basis of their X-ray to optical flux ratios only we run the risk of missing interesting sources. For instance, ULXs may display different values for log(F$_X$/F$_{\textrm{opt}}$) when observed in the high and low states, as was shown for M101 ULX-1 and M81 ULS1 (\citealt{tao11}). For both sources log(F$_X$/F$_{\textrm{opt}}$) is between 2 and 3 during the high state, but around 0 during the low state, well inside the range of optical to X-ray flux ratios found for AGN. Therefore other source properties should be taken into account as well, such as galaxy morphology, the distance of the ULX to its apparent host galaxy and the absolute magnitude of its optical counterpart. 
The source in \am{} is a good example of a candidate with such favorable properties: situated in a ring galaxy, which is a strong sign of a recent interaction phase that triggered star formation, often linked to ULXs (e.g. \citealt{swartz04}), and close to the center of its apparent host galaxy, decreasing the chance that it is a background AGN (\citealt{wong08}). It has an optical counterpart of such magnitude that it is consistent with being a bright globular cluster if it is at the distance of \am{}. Nevertheless our optical spectrum showed it to be a background object.

\subsection*{The ULX in \eso}
The X-ray source in \eso~is the only one of the four candidates in our sample that appears to be a bona fide ULX.  The extended nature of the source is confirmed by the fact that the emission line spectrum is consistent with that of an HII region. However, the [OI]/H$\alpha$ ratio indicates that some of the ionizing flux could come from an X-ray source. Potentially, we have found a ULX embedded in an HII region. 
Another possibility is that this ULX candidate is a background AGN shining through an HII region in \eso. The X-ray to optical flux ratio is similar to that of the other AGN in our sample, so we would expect to see a contribution of redshifted emission lines from the AGN in the optical spectrum. The fact that we do not detect this makes this scenario implausible. 

We find a 2-$\sigma$ upper limit for the flux of a  HeII $\lambda4686$ line of $10^{-17}$ \flx. This corresponds to an upper limit to the luminosity in the line of $\sim 5 \times 10^{36}$ \lum. The presence of this line would be a strong indication of ionization by an X-ray source. We can compare this upper limit with the strength of the HeII $\lambda4686$ line in other ULX nebulae. For Holmberg II X-1, \citet{pakull02} report a luminosity of $2.5 \times 10^{36}$ \lum. \citet{kaaret09} report a flux for this line from the ULX in NGC 5408 of $3.3 \times 10^{-16}$ \flx, which translates to a luminosity of $9 \times 10^{35}$ \lum~at the distance to NGC 5408 (4.8 Mpc, \citealt{karachentsev02}). Both these ULXs have an X-ray luminosity of $\sim 10^{40}$ \lum -- similar to \eso~-- and a HeII $\lambda4686$ to X-ray luminosity ratio of $\sim 10^{-4}$. If the same is true for \eso~then we would expect a HeII $\lambda4686$ flux of a few times $10^{-18}$ \flx, which is just below our 2-$\sigma$ upper limit. New observations of this source with greater sensitivity at the wavelength of the HeII $\lambda4686$ line are needed to determine if the nebula is X-ray photo-ionized or not.

\section*{Acknowledgements}
PGJ and MAPT acknowledge support from the Netherlands Organisation for Scientific Research. GM acknowledges support from the Spanish Plan Nacional de Astronom\'{\i}a y Astrof\'{\i}sica under grant AYA2010-21490-C02-02. This research is based on observations made with ESO Telescopes at the La Silla Paranal Observatory under programme ID 088B-0076A. This research has made use of software provided by the Chandra X-ray Center (CXC) in the application package CIAO and of the software packages Pamela and Molly provided by Tom Marsh.

 \bibliographystyle{mn_new}
 \bibliography{bibliography}

\end{document}